# Perovskite Substrates Boost the Thermopower of Cobaltate Thin Films at High Temperatures

P. Yordanov[1], P. Wochner[1], S. Ibrahimkutty[1], C. Dietl[1], F. Wrobel[1], R. Felici[2], G. Gregori[1], J. Maier[1], B. Keimer[1] and H.-U. Habermeier[1]

(1) Max Planck Institute for Solid State Research, 70569 Stuttgart, Germany

(2) Instituto SPIN-CNR, 00133 Roma, Italy

**Abstract**

Transition metal oxides are promising candidates for thermoelectric applications, because they are stable at high temperature and because strong electronic correlations can generate large Seebeck coefficients, but their thermoelectric power factors are limited by the low electrical conductivity. We report transport measurements on $Ca_3Co_4O_9$ films on various perovskite substrates and show that reversible incorporation of oxygen into $SrTiO_3$ and $LaAlO_3$ substrates activates a parallel conduction channel for p-type carriers, greatly enhancing the thermoelectric performance of the film-substrate system at temperatures above 450 °C. Thin-film structures that take advantage of both electronic correlations and the high oxygen mobility of transition metal oxides thus open up new perspectives for thermopower generation at high temperature.

The search for new thermoelectric materials has focused on the figure-of-merit $ZT = PT/\kappa$, where $P = S^2\sigma$ is the power factor, $S$ the Seebeck coefficient, $\sigma$ the electrical conductivity, $T$ the temperature, and $\kappa$ the thermal conductivity. $ZT$ values exceeding unity, which are desirable for energy harvesting applications, have thus far been realized mainly in tellurium-, antimony-, and germanium-based compounds [1][2]. However, most of these materials are not stable enough to endure the high-temperature conditions required for waste-heat recovery in environments such as combustion engines, coal power plants, and foundries. This situation has motivated research on the thermoelectric properties of cobalt oxides, which exhibit large Seebeck coefficients due to spin and orbital correlations in the cobalt $d$-electron system, and are environmentally benign and chemically stable at high temperatures. The most widely investigated compound is the layered cobaltate $Ca_3Co_4O_9$ (CCO), which shows a positive Seebeck coefficient of $S \sim 200$ µV/K and is stable up to at least 800 °C. Due to its rather low electrical conductivity, however, the $ZT$-values of CCO and its derivatives have remained below 0.5, despite various efforts to enhance the thermoelectric performance by chemical substitution [3][4].

To explore the potential of thin-film technology in optimizing the thermoelectric performance, we have epitaxially deposited CCO films of varying thickness on different oxide substrates and studied their Seebeck coefficient and electrical conductivity. For films with thickness below 50 nm on perovskite-type $SrTiO_3$ (STO) or $LaAlO_3$ (LAO) substrates, our data reveal a large (up to a factor of 6) enhancement of the thermopower for $T \geq 450$ °C. Experiments in controlled atmospheres show that this enhancement is due to a reversible oxygen incorporation reaction which activates $p$-type electrical conductivity in STO and LAO at high temperatures. These results demonstrate a functional decoupling of the electrical transport properties of CCO films from their effective thermopower, and suggest that perovskite substrates can serve as power-factor amplifiers.

Epitaxial CCO films were synthesized by pulsed laser deposition with a KrF excimer laser (Coherent Inc., wavelength 248 nm) using a repetition rate of 3 Hz, background oxygen pressure of 0.1 mbar, and substrate temperature of 650 °C [5]. The phase purity, film thickness, and surface morphology were determined by x-ray diffraction and atomic force microscopy (see the supplementary materials). High-quality films could be grown on STO, LAO, $[LaAlO_3]_{0.3}$-$[Sr_2AlTaO_6]_{0.7}$ (LSAT), $LaSrAlO_4$ (LSAO), MgO, and $Al_2O_3$, showing that CCO can accommodate a large degree of strain, most likely due to its misfit lattice structure [6]. The temperature dependent Seebeck coefficient $S(T)$ and electrical resistance $R(T)$ were measured simultaneously in 950 mbar oxygen, using a ZEM3 system (ULVAC) equipped with Pt electrodes. Each $S(T)$ data point was evaluated by three consecutive measurements with temperature differences $\Delta T$ = 20, 30, and 40 °C.

Figure 1a shows the temperature dependent Seebeck coefficient of a series of CCO films. Films on LSAT, LSAO, and MgO follow the behavior of bulk CCO with a *T*-linear increase up to the highest measured temperature, 720 °C, where *S* reaches values of 170-200 µV/K in agreement with prior work [7]. Remarkably, however, films on STO and LAO show a strong enhancement of *S* upon heating above 450 °C. The upturn of *S(T)* is most pronounced for films with thickness $d_{film}$ ≤ 50 nm where the maximal *S*(720 °C) reaches 800 µV/K and even exceeds 1000 µV/K for STO and LAO substrates, respectively. Since control experiments using other materials deposited on STO showed qualitatively similar effects (see the supplementary material), we can rule out that the enhancement of the Seebeck coefficient is an intrinsic property of the CCO films. Rather, the results indicate a contribution of the substrate material and/or the substrate-film interface. The positive sign of this contribution implies positively charged mobile carriers.

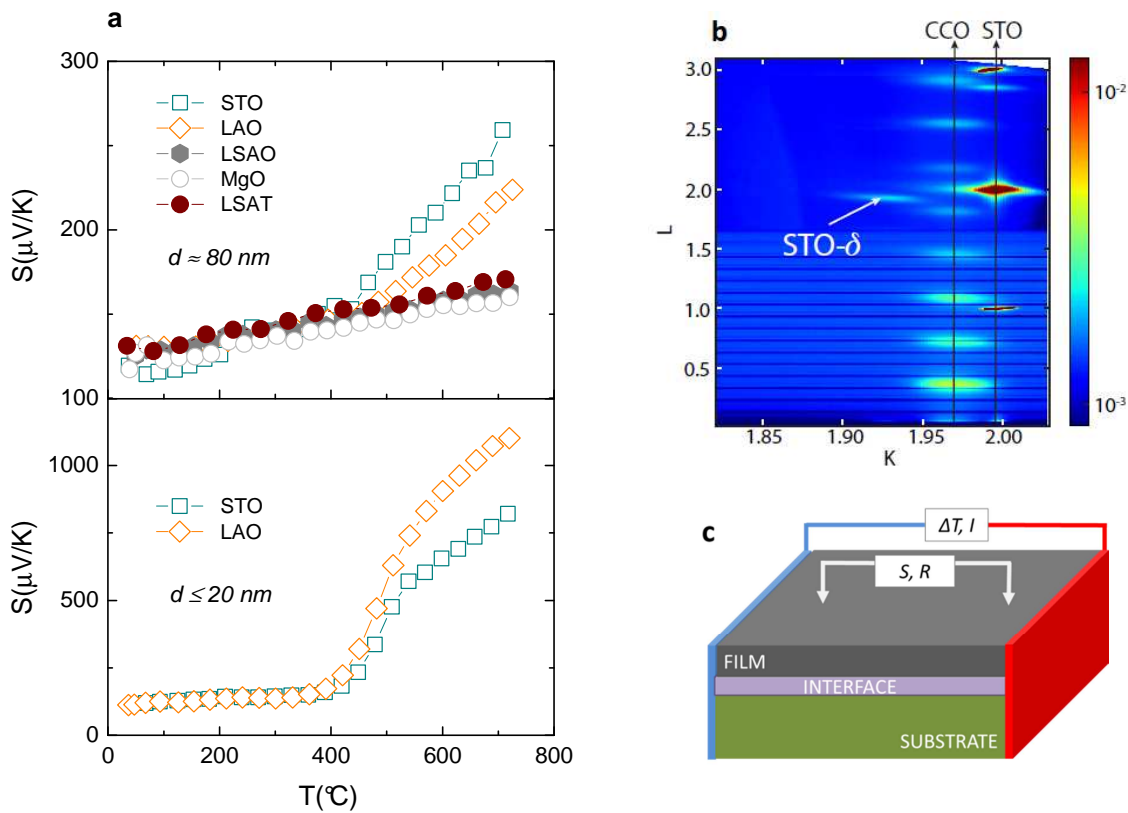

FIG. 1 (Color online). (a) Temperature dependent Seebeck coefficient of CCO films with thickness 80 nm (upper panel) and 20 nm (lower panel) deposited on various oxide substrates. The measurements were performed in 950 mbar oxygen and were stable upon temperature cycling. (b) Reciprocal space map of an 80 nm thick film on STO, carried out at beamline ID03 of the European Synchrotron Radiation Facility (ESRF) using photon energies of 20 keV. Axes are labeled in STO lattice units. The map shows the (2KL) plane with the (22L) reflections of pure

STO and oxygen deficient STO-δ, and the (02L) reflections of the $CoO_2$ subsystem of the CCO lattice. Weak intensity from the (404) reflection of the interfacial rock-salt phase $Ca_2CoO_3$ is seen near the (023) reflection of STO. (c) Experimental setup for measurements of the Seebeck coefficient $S$ and electrical resistance $R$.

To elucidate possible contributions of the substrate and the substrate-film interface to the thermoelectric properties, we carried out detailed synchrotron-based high-resolution x-ray diffraction experiments on selected thin-film structures. Figure 1b shows the reciprocal-space map of an 80 nm thick CCO film grown on STO. In addition to the Bragg reflections from the CCO film and the STO substrate, we found reflections indicative of off-stoichiometric STO with an expanded lattice parameter of 4.05 Å at room temperature. This effect could arise from oxygen vacancies near the substrate-film interface; a comparison to data on bulk $SrTiO_{3-\delta}$ yields $\delta \sim 0.5$ [8]. Likely, however, other factors such as deviations of the Sr/Ti ratio from stoichiometry and/or intermixing with CCO components during film growth also contribute to the lattice expansion [9]. In any case, the thickness, $d_{int}$, of the interfacial layer extracted from the widths of the corresponding $SrTiO_{3-\delta}$ Bragg peaks was found to be 7-10 nm for several CCO films on STO. Oxygen depletion regions of comparable thickness were also observed for films on LSAT and LAO substrates, although not as well-defined in the latter case.

Motivated by the observation of oxygen-deficient interface layers, we monitored the response of the thermoelectric properties of our thin-film structures to changes in the oxygen content of the surrounding atmosphere. Figure 2a shows the resistivity and Seebeck coefficient of a CCO film on STO at $T \sim 600$ °C (i.e., well within the thermopower enhancement regime) as the oxygen partial pressure is modified from 100% to 1% and back. The reversible decrease of $S$ upon lowering the oxygen concentration indicates that oxygen is incorporated into the system and plays a key role in the thermopower enhancement. We note that such a partial pressure influence on the thermopower is just the opposite compared to that of bulk CCO [10]. This finding confirms that the additional contribution originates from the substrate and/or substrate-film interface. STO and LAO are indeed known for their low oxygen vacancy migration barriers (~ 0.6 eV according to theoretical estimates [11][12]). The much lower oxygen mobility and higher resistivity in other substrate materials such as LSAT (see supplementary material) then explains why the thermoelectric properties of these films are not affected by oxygen incorporation, despite the presence of oxygen-deficient interface layers revealed by x-ray diffraction.

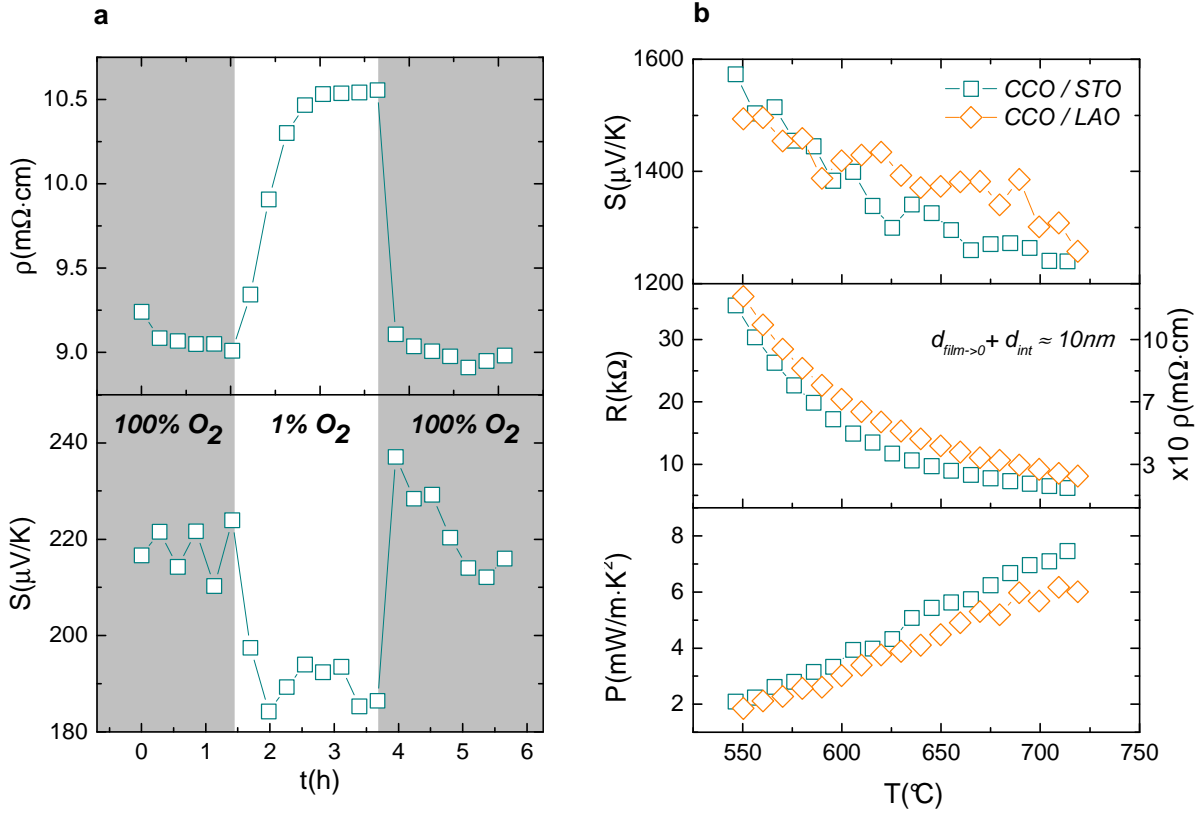

FIG. 2 (a) Oxygen partial pressure dependence of the resistivity (upper panel) and Seebeck coefficient (lower panel) of a CCO film with $d_{film}$ = 80 nm deposited on STO. Grey shaded areas correspond to *100 %* O$_2$, and the unshaded area to a mixture of *1 %* O$_2$ and *99 %* N$_2$. The measurements were performed in a narrow temperature range *T* = 570-600 °C, at constant total pressure of 950 mbar. (b) Temperature dependent Seebeck coefficient *S*, resistance *R*, resistivity *ρ*, and power factor *P* for ultrathin CCO films prepared by 180 laser pulses, corresponding nominally to $d_{film}$ ~ *2-3 nm*, on STO and LAO substrates. *ρ* and *P* were estimated by assuming that the current flows in the topmost layer of thickness $d_{film} + d_{int}$ ~ 10 nm.

These results suggest that oxygen diffusion into the oxygen-deficient interface layers of STO and LAO is responsible for the thermopower enhancement at temperatures exceeding 450 °C (Fig. 1). Oxygen atoms occupying vacancies in these layers are expected to release two mobile holes which open a conduction channel parallel to the CCO film. In the framework of a slab model, the effective resistance and Seebeck coefficient of the resulting parallel-conduction network can then be expressed as $1/R_{total} = \sum_i 1/R_i$ and $S_{total} = \sum_i (S_i \sigma_i d_i)/\sum_i (\sigma_i d_i)$, respectively, where *i = film, int, sub* denote the three subsystems (film, interface, and substrate) [13][14]. Depending on the contributions of the individual components, the effective Seebeck coefficient of the film-substrate system can be larger than the one of the CCO film alone. In view of the large Seebeck coefficients of bulk substrate materials at high temperatures (e.g. ~1200 *µV/K* for STO), [15][16] it is conceivable that the contribution of the off-stoichiometric interface layer substantially enhances $S_{total}$, once the conductivity from carriers generated through oxygen incorporation becomes large enough.

To explore whether this mechanism is indeed responsible for the upturn of the thermopower at high temperatures (Fig. 1), we estimated the contributions of the individual components in the slab model through a series of experiments. A comparison to extensive data on CCO films on LSAT, LSAO, and MgO (Fig. 1) justifies a linear extrapolation of the low-temperature behavior to estimate $S_{film}(T)$ and $\sigma_{film}(T)$ over the entire temperature range up to 720 °C. $S_{sub}(T)$ and $\sigma_{sub}(T)$ were determined through measurements on samples from which the CCO films were removed by polishing. The results agree with literature values (see the supplementary materials). Estimating the contribution of the interface layer is somewhat more difficult, because its precise composition and oxygen vacancy concentration are established during the growth of the CCO film. We therefore measured *S(T)* and *R(T)* on ultrathin films on STO and LAO. Figure 2b represents the experimental data and the estimated *ρ(T)* and *P(T)* assuming that the conductivity is entirely due to the topmost layer of thickness $d_{film} + d_{int}$ ~ 10 nm determined by x-ray diffraction. The S(T), however, is influenced by the bulk values of STO, LAO substrates, respectively. An estimate of the interface Seebeck coefficient in the Heikes model, corresponding to *ρ(T)* ≈ *20 mΩ·cm*, mobility *µ*≤1 cm$^2$/V·s, relaxation time *τ* ~10$^{-15}$ s and carrier density of *p*~10$^{20}$ cm$^{-3}$ would be $S_{int}$ ~ 450-500 µV/K. As a final cross-check, we repeated one of the measurements after reducing the substrate thickness by polishing from the back side (see the supplementary material). The results confirm the validity of the parallel slab model to $S_{total}$ and $R_{total}$ with $R_{film} << R_{sub}$.

With parameters determined in this way and $S_{int}$ as a fit variable, the slab model yields excellent descriptions of the Seebeck coefficient and resistivity of thicker CCO films. As an example, Fig. 3a shows that the model quantitatively describes the upturn of *S(T)* at 450 °C for a 27 nm thick film on STO with $S_{int}(fit)$=540 µV/K, consistent with the estimate given above. The upturn is explained as a consequence of the increase of $\sigma_{int}(T)$ and $\sigma_{sub}(T)$ by oxygen incorporation, such that the products $S_{film}d_{film}\sigma_{film}$, $S_{int}d_{int}\sigma_{int}$ and $S_{sub}d_{sub}\sigma_{sub}$ become comparable.

Whereas the interface characteristics and its estimated contribution to $S_{total}$ contain considerable uncertainties, we emphasize its functional importance to the thermopower enhancement mechanism. It is very likely that the substrate bulk acts as a reservoir for oxygen vacancies while the oxygen incorporation takes place predominantly at the interface layer on time scale comparable with the measurements.

In a similar manner, the model also quantitatively describes *S(T)* data for thin-film structures of different materials such as Pt on STO (see the supplementary material). It also explains why the thermopower enhancement is only observed for CCO films with $d_{film} \leq 50$ nm, as $S_{total}$ for thicker films is dominated by the film contribution (Fig. 3b). Open issues to be explored by future research include the precise chemical composition of the interface layers, the possible presence of a space charge layer at the interface, and the question whether ionic transport contributes to the conductance channel opened by oxygen incorporation. The activation energies of 0.4-0.5 eV extracted from the experimentally measured transport data (Fig. 2b) favor an electronic origin, but do not exclude an ionic contribution.

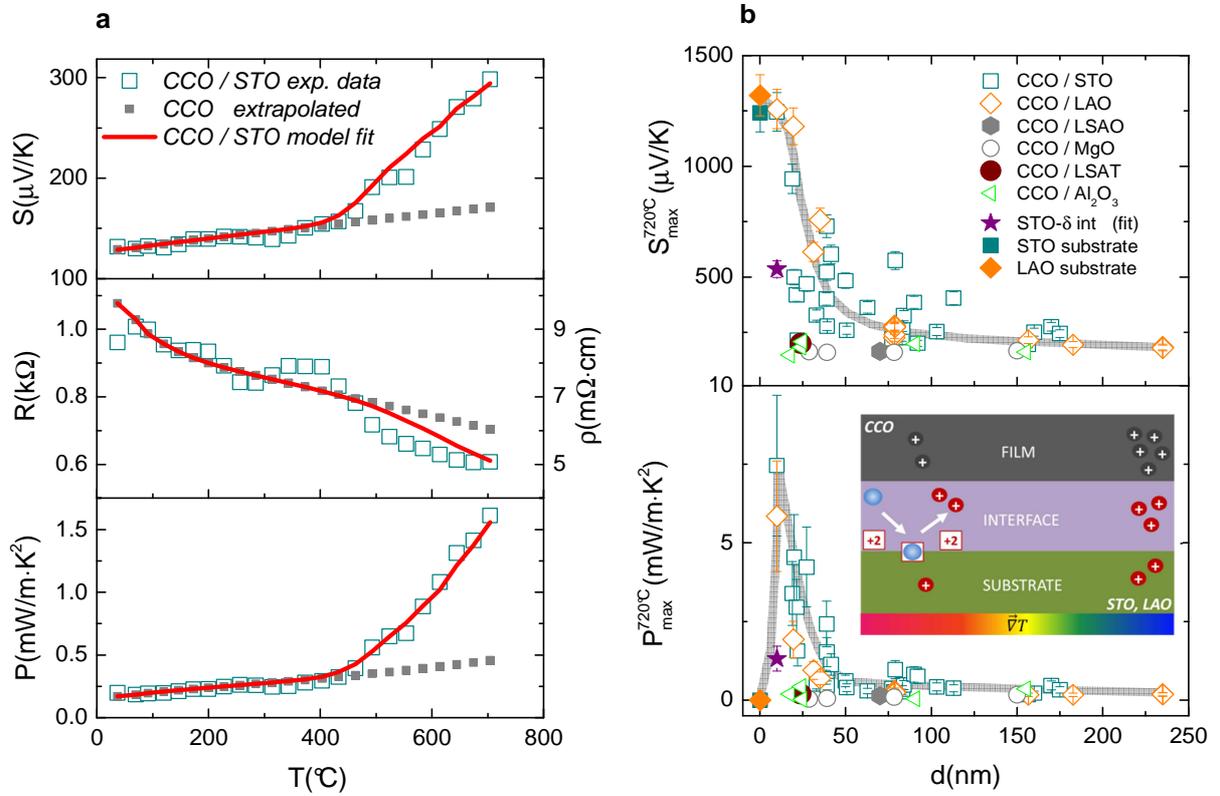

FIG. 3 (a) Data for a CCO film with $d_{film}$ = 27 nm on STO. Red solid lines represent the results of a model calculations described in the text, and filled grey squares the extrapolated low-temperature data of CCO. The input parameters at 720 °C read; the CCO film: $S_{film}$=170 µV/K, $R_{film}$=0.7 kΩ, $d_{film}$=27 nm, $\rho_{film}$=6.4 mΩ·cm; for the interface layer: $S_{int}$(fit)=540 µV/K, $R_{int}$=6.5 kΩ, $d_{int}$=10 nm, $\rho_{int}$=21.75 mΩ·cm; and for the STO substrate: $S_{sub}$=1240 µV/K, $R_{sub}$=16 kΩ, $d_{sub}$=0.5 mm, $\rho_{sub}$=11.4 Ω·m. (b) Experimental Seebeck coefficient *S* (upper panel) and estimated power factor *P* (lower panel) at the highest measured temperature of *T* = 720 °C for CCO thin films deposited on different substrates as a function of the film thickness. $P$ was estimated by assuming that the current flows in the topmost layer of thickness $d_{film} + d_{int}$; the large error bar accounts for uncertainties in this estimate. Reference values for pristine STO and LAO substrates as well as the result from the model fit for the interfacial STO layer are also given. Solid lines are guides to the eye. The inset shows a sketch of the slab model described in the text.

In conclusion, we have shown that oxygen incorporation into the STO (LAO) interface / substrate system can substantially enhance the Seebeck coefficient of cobaltate thin films at high temperatures. This mechanism partially offsets the limited electrical conductivity of bulk cobaltates. Assuming electric currents exclusively due to the CCO films, the power factor is estimated in the *mW/m·K$^2$* range (Fig. 3b), which is quite competitive with other high-temperature thermoelectric materials. Our results demonstrate that the thermoelectric properties of transition metal oxide films at high temperatures can be substantially modified through the film thickness and choice of substrate, thus pointing out new perspectives in the search for high-performance thermoelectrics.

**Supplementary Material**

See Supplementary Material for additional information on the sample characterization by x-ray diffraction and atomic force microscopy, as well as further thermoelectric reference measurements.

**Acknowledgments**

We acknowledge financial support by the DFG (TRR 80) and by Coherent Inc., Wilsonville, OR (USA). The authors are indebted to I. Fritsch, R. Merkle, G. Logvenov, W. Xie, A. Weidenkaff, and S. Populoh for experimental support and discussions.